%% file: template.tex
\documentclass{article}

\usepackage{arxiv}

\usepackage[utf8]{inputenc}     
\usepackage[T1,T2A]{fontenc}    
\usepackage{hyperref}           
\usepackage{url}                
\usepackage{booktabs}           
\usepackage{amsfonts}           
\usepackage{nicefrac}           
\usepackage{microtype}          
\usepackage{lipsum}

\usepackage{graphicx}

\usepackage{amssymb}

\usepackage{lineno}

\usepackage{subcaption}

\usepackage{doi}

\usepackage{diagbox}

\usepackage{enumitem}

\usepackage[usenames, dvipsnames]{color}
\usepackage{xr}
\usepackage{comment}

\usepackage{textcomp}
\emergencystretch=\maxdimen
\title{PlugSonic: a web- and mobile-based platform for binaural audio and sonic narratives}

\author{
    Marco Comunit\`a\\
    Dyson School of Design Engineering\\
    Imperial College London\\
    United Kingdom\\
    \texttt{m.comunita@imperial.ac.uk}\\
    \And
    Andrea Gerino\\
    Dyson School of Design Engineering\\
    Imperial College London\\
    United Kingdom\\
    \texttt{a.gerino@imperial.ac.uk}\\
    \AND
    Veranika Lim\\
    Dyson School of Design Engineering\\
    Imperial College London\\
    United Kingdom\\
    \texttt{v.lim@imperial.ac.uk}\\
    \And
    Lorenzo Picinali\\
    Dyson School of Design Engineering\\
    Imperial College London\\
    United Kingdom\\
    \texttt{l.picinali@imperial.ac.uk}\\
}

\begin{document}
\maketitle

\begin{abstract}
PlugSonic is a suite of web- and mobile-based applications for the curation and experience of binaural interactive soundscapes and sonic narratives. It was developed as part of the PLUGGY EU project (Pluggable Social Platform for Heritage Awareness and Participation) and consists of two main applications: PlugSonic Sample, to edit and apply audio effects, and PlugSonic Soundscape, to create and experience binaural soundscapes. The audio processing within PlugSonic is based on the Web Audio API and the 3D Tune-In Toolkit, while the exploration of soundscapes in a physical space is obtained using Apple's ARKit. In this paper we present the design choices, the user involvement processes and the implementation details. The main goal of PlugSonic is technology democratisation; PlugSonic users - whether institutions or citizens - are all given the instruments needed to create, process and experience 3D soundscapes and sonic narrative; without the need for specific devices, external tools (software and/or hardware), specialised knowledge or custom development. The evaluation, which was conducted with inexperienced users on three tasks - creation, curation and experience - demonstrates how PlugSonic is indeed a simple, effective, yet powerful tool.
\end{abstract}


\section{INTRODUCTION}
\label{introduction}

A heritage that is everywhere, and is relevant to everyday life, is one of the preconditions for genuine sustainability \cite{Fairclough2014:Faro}. Currently, there are few ICT tools to support citizens in their everyday activities to shape cultural heritage and be shaped by it. Furthermore, existing applications and repositories for heritage dissemination do not foster the creation of heritage communities (e.g. Google Arts and Culture, Europeana) \cite{Russo2012:Rise} \cite{Stuedahl2012:Heritage}. Social platforms certainly offer potential to build networks, but they have not yet been fully exploited for global cultural heritage promotion and integration in people’s everyday life \cite{Lim2018:Pluggy} and museums and institutions have only recently started to explore the potential of social media and technology for public engagement and co-creation purposes \cite{Russo2008:Participatory}.

The PLUGGY project (Pluggable Social Platform for Heritage Awareness and Participation) \cite{PluggyProject},  aims to bridge this gap by providing the necessary tools to allow users to share their local knowledge and everyday experience with others and, together with the contribution of cultural institutions, to build extensive networks around a common area of interest, connecting the past, the present and the future.

\begin{figure}[t]
	\centering
	\includegraphics[width=.75\linewidth]{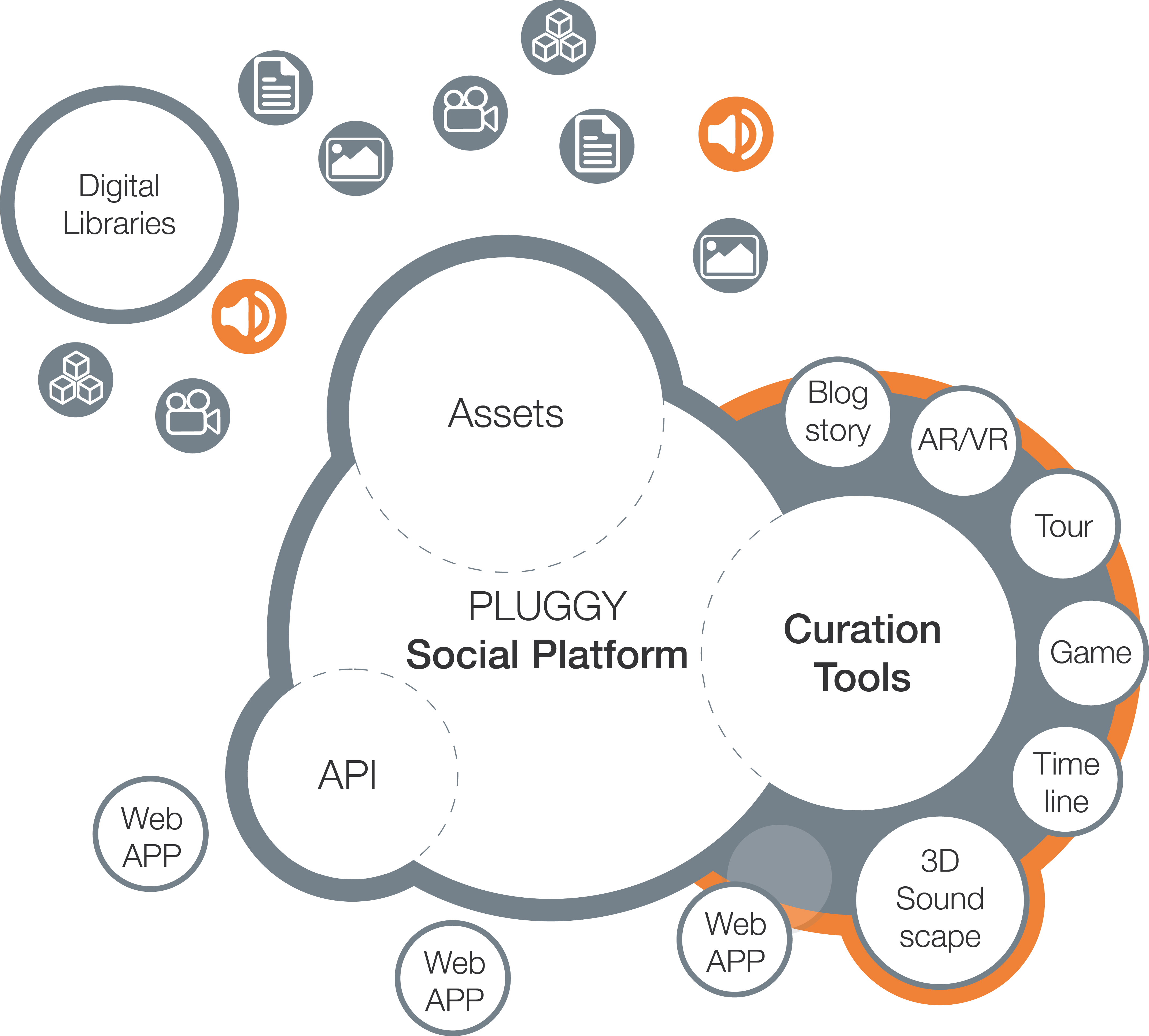}
	\caption{The PLUGGY social platform and apps structure}
    \label{fig:pluggy_diagram}
\end{figure}

Within PLUGGY, several tools are being developed (Figure \ref{fig:pluggy_diagram}): a social platform, curation tools, and 4 "pluggable" apps to demonstrate the platform's potential and kick-start applications for the after-project life.

The PLUGGY social platform serves multiple purposes. It is a repository for all the content uploaded (assets) and curated (exhibitions) by users and institutions, as well as a tool for interaction and collaboration. It has all the typical features of a social platform: profile, private and public content, follow, like, comment, share, notifications, folders to organise bookmarked content, teams for collective contribution. At the same time, the social platform gives access to the curation tools - which are designed to create basic exhibitions in the form of blog stories or time-lines - and is extended by the "pluggable" apps - for the curation of augmented exhibitions and engaging experiences. These applications focus on various aspects of digital heritage, which include Virtual (VR) and Augmented Reality (AR), Geolocation, Gamification and 3D Soundscapes and Sonic Narratives. The latter, called PlugSonic, is the focus of the current paper, which will look in particular at the web- and mobile-based binaural audio features of the application.

With its focus on open access and participation, the PLUGGY social platform gives also access to external repositories like Europeana \cite{Europeana} and Wikipedia and includes an open source application programming interface (API) which allows anyone to develop new applications that access the platform's content and extend its features and curation tools. Any user can request to become a developer and - after adequate evaluation and approval - "plug" a new app to the platform for everyone to benefit.

The research presented here aims at contributing to both the cultural heritage and spatial audio communities through the development of novel tools for the creation and experience of realistic and interactive 3D soundscapes. Special importance has been given in designing applications that allow anyone, with no knowledge or previous experience in 3D audio production or soundscapes curation, to contribute to the dissemination and growth of sonic cultural heritage. Being web- and mobile- based, the tools presented here are intrinsically ubiquitous and the content can be experienced online or in a physical space, without the need for specialised software or hardware installation, and/or custom development. We also conducted an evaluation with inexperienced users to substantiate our objective towards a very low barrier to the content development, while retaining the effectiveness in public engagement of spatial audio technologies.

The paper is organised as follows: section 2 provides a background about binaural spatialisation, as well as a review of the state of the art for web-based spatial audio and the use of sonic narratives in cultural heritage. Section 3 describes all the apps (web- and/or mobile-based) that are part of the PlugSonic \textit{suite}, main functionalities (complete details are included in the appendices) and implementation choices. In section 4 we report on the evaluation methods and discuss the results. Section 5 is devoted to the conclusions and potential future work.

\section{BACKGROUND}
\label{background}
Sonic narratives are generally based on music features (e.g. timbre, pitch-melody, tempo, etc.) \cite{Delle2012:Paper}, and are often not interactive (i.e. simple audio playback). The addition of spatial attributes (e.g. placement of sound sources on a full 360° sphere, and at different distances), and most of all the addition of interactivity (e.g. to navigate soundscapes moving around in the acoustic virtual environment), are features which have not been widely explored until now - for example, \cite{Krakowsky} explored spatial sonic narratives, but exploited simple 2-dimensional audio panning techniques. The idea of developing and evaluating web- and mobile- applications for the creation and experience of 3D sonic narratives does indeed represent a novel contribution to both the digital heritage and audio technology domains.

\subsection{BINAURAL SPATIALISATION}

The aim of binaural spatialisation is to provide the listener (through standard headphones) with the impression that sound is positioned in a specific location in the three-dimensional space. The 3D characteristics of the sound can be captured during recording with special hardware, or simulated in post-production via spatialisation techniques.
The theories at the basis of the binaural spatialisation technique are not particularly recent, and the first binaural recording dates back to the end of the 19th Century \cite{Collins2008:Theatrophone}. However, it is only within the last twenty years that the increase in the calculation power of personal computers enabled an accurate real-time simulation of three-dimensional sound-fields over headphones.

Several tools currently exist for performing binaural spatialisation; just to mention a few, Anaglyph \cite{Poirier2018:Anaglyph}, IRCAM Spat \cite{Carpentier2015:Twenty}, the IEM binaural audio open library \cite{Musil2005:Library}, and the 3D Tune-In Toolkit \cite{Cuevas2019:3d}. Very few of these are though implemented within a web-based application, and therefore available on multiple platforms through a simple browser.

\subsection{WEB-BASED SPATIAL AUDIO}

With the specification and release of the Web Audio (WAA) \cite{WebAudioApi} and the Web GL (WGL) \cite{WebGL} APIs in 2011, and with the later release of the HTML5 specification in 2014; the World Wide Web Consortium (W3C) and the Mozilla Foundation set the basis for the development of modern web applications.
As stated in the introduction to the WAA \cite{WAASpec}, the specification of a high-level Javascript (JS) API was necessary to satisfy the demand for audio and video processing capabilities to implement "sophisticated web-based games or interactive applications". To obtain performances comparable to modern digital audio workstations (DAWs) and games graphical engines, the decision was taken to move the burden of audio and video processing from the server to the client (i.e. browser) side.

The WAA inherited the idea of modular routing, where complex signal processing chains are created connecting many audio blocks. Built around the \textit{AudioNode} object; this class defines audio blocks that are sources, destinations or processing units having inputs, outputs or both.
The API also allows for the spatialisation of sound through an equal power panning and an HRTF(Head Related Transfer Function)-based convolution algorithms.
For headphones based applications, the equal power panning algorithm “can only give the impression of sounds located between the ears” \cite{Pike2015:Delivering} while, as reported in \cite{Carpentier2015:Binaural}, the HRTF algorithm implemented in Google Chrome and Mozilla Firefox embeds only one set of HRTFs from the IRCAM Listen database \cite{IrcamListen}.
Carpentier \cite{Carpentier2015:Binaural} discusses the limitations and potential drawbacks for users and developers of having only one choice of HRTFs (e.g. in-head localisation, inaccurate lateralisation, poor elevation perception and front-back confusion) and presents novel work to implement a \textit{BinauralFIRNode} class. This class extends the WAA allowing to import custom HRTFs as FIR filters.

Another limitation of the WAA is in the method used for the simulation of distance; based on attenuation only, it does not account for frequency domain effects of waves propagation.

Even with its limitations the WAA is having a great impact on web-based applications and research on spatial audio, audio narratives, games and immersive content broadcasting.
Pike et al. \cite{Pike2015:Delivering} developed an object-based and binaural rendering player with head-tracking while RadioFrance nouvOson website \cite{Dejardin2015:Nouvoson} broadcasts 5.1 surround and binaural audio.

The WAA has also been used to develop high-order Ambisonics sound processing. Google Omnitone \cite{Omnitone} implements decoding and binaural rendering up to the third order (16 channels) and uses eight virtual loudspeakers based on HRTF convolution to render binaural audio.
This library is also used in the Google Resonance Audio SDK \cite{Resonance}, which extends it with: sound source directivity customisation, near-field effects, sound source spread, geometry-based reverb (room dimensions and wall materials) and occlusions.
A relevant application of Omnitone in the field of soundscapes is the Storyspheres web app by the Google News Lab \cite{Storyspheres} which allows to create a soundscape for a $360^\circ$ image placing sound sources in the scene and render the spatialised sound.

In \cite{Politis2016:JSambisonics}, Politis and Porier-Quinot present JSAmbisonics as a library that uses the WAA for interactive spatial sound processing on the web. This work is interesting for it supports Ambisonics of any order.

Great effort is also being put into investigating and developing web-based soundscape creation and experience tools with features and capabilities comparable to plug-ins that are used within DAWs by professionals to create immersive audio content. This trend exemplifies the interest and the aim to make 3D technologies open, familiar and available to a broad public.

The INVISO project \cite{Camci2017:Inviso} focused on the development of a web-based interface for “designing and experiencing rich and dynamic sonic virtual realities” suitable for both experts and novices. Here, the WGL was used to design a 3D interface for the creation of sound objects and navigation of the environment and the WAA for the audio rendering. The project presents interesting features like the multi-cone sound object to model complex directivity patterns; the control over sources’ elevation; and the possibility to define trajectories for moving sources or sound zones in which sounds are not spatialised to create ambient sounds. Nevertheless, as discussed in previous paragraphs, the rendering quality was limited by the use of the WAA and the app did not offer control over other important aspects of an interactive soundscape, such as the playback order or the areas within which sound sources are audible. Differently from the work presented here, which includes a mobile app, within INVISO it was not possible to experience the soundscapes in real environments. Furthermore, the evaluation was conducted mostly (9 out of 10 cases) with participants with previous experience with DAWs or VR authoring software. On the other hand, the development of PlugSonic within the framework of a social platform aimed specifically at encouraging naive users to create and share new content, allowing them to store and access it from anywhere on any device.

Another interesting concept, even if it does not use spatialisation, is the SimScene tool by Rossignol et al. \cite{Rossignol2015:Simscene}. Here the idea is to create soundscapes but without allowing the user complete control over the sound scene, in this way adding a degree of randomisation to the soundscape. As described in the paper, users do not select or interact with individual sounds, but with group of sounds belonging to the same class; for each class they only have control over high level parameters (sound intensity, spacing time average and variance). The soundscape is represented on a timeline where the user places blocks of sound classes with control over the item and global fade-in/fade-out time.

To conclude this paragraph, it is also worth citing other APIs available for the integration into web apps. Facebook recently released (July 2018) a JS implementation of their Audio360 \cite{Audio360} renderer to include Ambisonics decoding in the web to be used to play 
videos created using their Spatial Audio Workstation plug-ins.

It is evident from the literature collected above that no tool currently exists for naive users (i.e. with no or little experience in audio engineering and/or music technology) to create interactive high quality binaural audio (e.g. with distance simulation, HRTF choice, etc.) on a simple-to-use web-based interface.

\subsection{SONIC NARRATIVES FOR CULTURAL HERITAGE}

In \cite{Ardissono2012:Personalization}, Ardissono et al. give an exhaustive review and comparative analysis of research about digital storytelling and the delivery of multi-media content, on-site or on-line, with a focus on cultural heritage. Here we will limit to those projects that use exclusively or mainly audio as medium for content delivery or to design novel types of experiences.

Looking at the state of the art in this area, it can be noticed how research is delving into solutions to make cultural heritage immersive, adopting augmented reality (AR), virtual reality (VR) and spatial audio; engaging, using personalisation and emotional storytelling; adaptive, exploiting context-awareness and location-awareness; interactive, using the paradigm of dramas; or open and inclusive, developing content for people with impairments and/or difficulties.

Two of the first projects aiming at enriching the user experience and engagement in a museum’s visit were the HyperAudio and the HIPS projects. In \cite{Not1998:Content} the authors introduce the HyperAudio project, in which the user is presented with audio content played through headphones when approaching an artefact, together with suggestions for further exploration on the display of a palmtop. In \cite{Petrelli2005:User} the hardware and software architecture is presented, which uses infrared transmitters nearby the exhibits and a receiver on the user’s headphones to implement a degree of location-awareness. The HIPS project \cite{Benelli1999:Hips} expands on the HyperAudio experience aiming at a user-adaptive presentation system.

The LISTEN project \cite{Zimmermann2008:Listen} investigated audio augmented environments and user adaptation technologies. This involved the development of ListenSpace \cite{Delerue2002:Authoring} - graphical authoring tool used to represent the real space and the sound sources' position - as well as the implementation of a domain ontology \cite{Zimmermann2003:Listen} for an exhibition and the use of context-awareness to adapt to the user's interest. For the spatial rendering the authors relied on IRCAM's Spatialisateur \cite{IrcamSpat}. The main limitations, from the content creators' perspective, could be seen into the system's software (server-based processing) and hardware (antennas or infrared cameras for head-tracking) requirements; and the necessity for custom development for each exhibition, together with the sound rendering only for the horizontal plane.

In the CHESS project \cite{Vayanou2014:Authoring} the focus was on personalisation and interactivity; where the story was delivered mainly through voice narration, with on-screen instructions and interaction obtained through applications running on web browsers. Personalisation was obtained through the definition of personae, profiling first-time visitors \cite{Pujol2013:Personalization} to adapt the narration style. Once profiled, the adaptive storytelling engine \cite{Vayanou2014:Authoring} used contextual data to adapt the user’s experience.

Interaction and context-awareness (based on geolocation) was explored in \cite{Hansen2012:Mobile} with mobile urban dramas (in which the user becomes the main character of a story). The project used a multimedia style (audio, video, images, animations) and was implemented to run on mobile web browsers using XML to describe the content. Here, the advantage of multi-platform flexibility, was limited by the need for specific knowledge about the content metadata structure or the consultancy from the researchers for app implementation and web services.

Other projects have been studying and developing new ways to attract visitors with engaging experiences. The EMOTIVE project \cite{Emotive} aimed to “research, design, develop and evaluate methods and tools that can support the cultural and creative industries in creating narratives” that exploit emotional storytelling. A first evaluation of a mobile application for the ancient Athens Agora \cite{Roussou2017:Engaging} produced positive results with the users particularly appreciating the chance of freely exploring the environment.

In the ARCHES project \cite{Arches} researchers worked on the inclusivity aspect of storytelling exploring new modalities to design cultural heritage experiences for people with difficulties and/or disabilities.
Other examples of tools developed by academia that could be used to create soundscapes and immersive experiences are the Soundscape Renderer \cite{Geier2012:Spatial} and those developed for the CARROUSO project \cite{Vaananen2003:User}.

There are also commercial applications developed for audio narratives like Podwalk \cite{Podwalk} and Echoes.XYZ \cite{Echoes}; both are GPS-triggered audio tours with web-based tools for the creation of audio guides and mobile apps to experience them. It is worth highlighting how none of the currently available commercial solutions use sound spatialisation.

\subsection{AIMS}
The overall aims of our research are illustrated below:
\begin{itemize}
    \item Develop tools that foster spatial audio for the creation of interactive soundscapes and audio narratives, with a focus towards cultural heritage
    \item Adopt web and mobile technologies to simplify and streamline the curation process for both the general public and institutions, while retaining a high quality audio rendering
    \item Prevent any need for specialised software, custom applications development and/or hardware requirements
    \item Prove that 3D audio technologies can be used by subjects without specialised knowledge and become part of our everyday life
    \item Conduct a thorough evaluation of the tools with inexperienced participants and assess whether: the tools are easy to understand and use, effective in the creation of virtual soundscapes for real environments, capable of delivering an engaging and compelling experience.
\end{itemize}

\section{PLUGSONIC SUITE}

In this section we describe the design criteria and implementation details of the web and mobile applications that make up the \textit{PlugSonic suite}. The apps - organised into the PlugSonic Sample and PlugSonic Soundscape groups - were developed to (1) facilitate the use of audio content to augment virtual exhibitions; (2) enhance on-line and/or on-site visits to museums, monuments, archaeological sites; (3) share tangible and intangible cultural heritage. Specifically, we designed PlugSonic Sample to edit sound files and apply audio effects, and PlugSonic Soundscape to create and experience interactive spatialised soundscapes. In this way, social platform, curation tools and pluggable apps can include standard sound files (i.e. mono/stereo, mp3/wav format) to be used in voice descriptions, audio narratives or sound accompaniment to the platform’s exhibitions, as well as interactive and explorable 3D audio narratives and soundscapes. PLUGGY users - whether institutions or citizens - are therefore given all the necessary instruments, and are not in need for specific devices, external tools (software and/or hardware), specialised knowledge and/or resources.

\begin{figure}
	\centering
	\includegraphics[width=.75\linewidth]{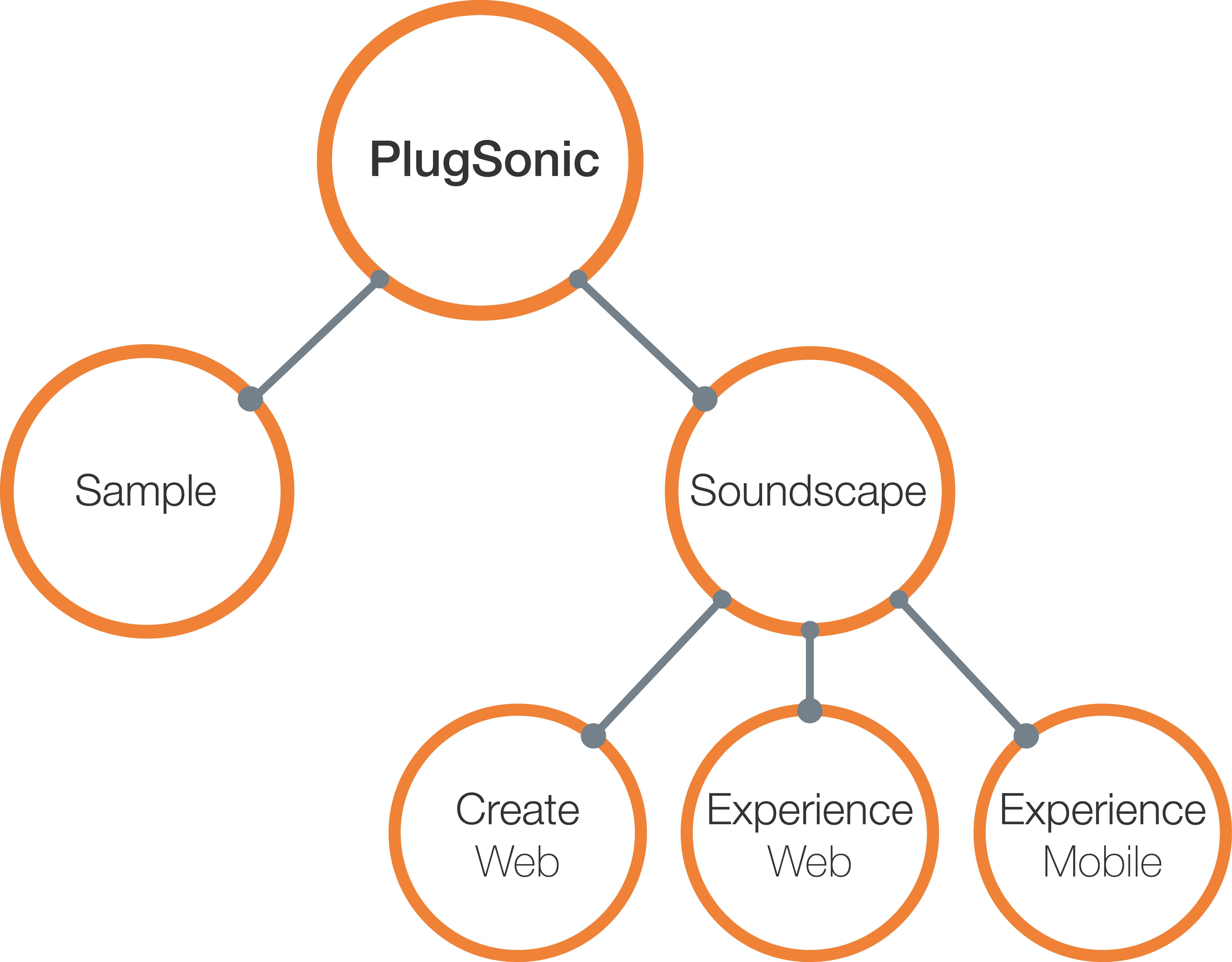}
	\caption{Apps of the PlugSonic suite}
    \label{fig:apps_diagram}
\end{figure}

Figure \ref{fig:apps_diagram} gives a complete overview of the structure of the web and mobile apps included in the PlugSonic suite. PlugSonic Sample and PlugSonic Soundscape Create were implemented only as web applications while PlugSonic Experience was implemented both as web and mobile application. In this way, all the soundscape exhibitions in PLUGGY can be explored in a virtual or physical space. When using Soundscape Experience Web, the navigation takes place on the browser using mouse, arrow keys or touch controls. With Soundscape Experience Mobile the navigation can take place in a real space; thanks to the use of the Apple ARKit \cite{ARKit} the user can freely navigate within a room while the device's camera is used to extrapolate the person's position in the room and hence in the soundscape.

A typical use case of the PlugSonic suite could be described as follows. Let's suppose that a user would like to curate a new 3D audio soundscape. The first step would be to select the appropriate audio assets to be included in the soundscape exhibition. The user has the option to either upload new audio files to the PLUGGY social platform or use content already available. Any file uploaded to PLUGGY (e.g. image, video, 3D model, sound) is associated to an asset - with a dedicated web page - which contains extended information such as: title, description, location, license, tags, number of likes, comments etc. While creating a new audio asset, the user might decide to use PlugSonic Sample to edit the audio file and enhance the sound. The next step would be to create the new soundscape exhibition using PlugSonic Soundscape Create. Upon loading the app, the user can set size and shape of the virtual environment (which might relate to a real physical environment or not) and search and retrieve the desired audio files. The sound sources are imported into the soundscape and represented as \textit{circles} in the virtual room. The curation process includes setting the position for each sound source together with many other options controlling the sources' sound and the interaction between sources and user (see section \ref{sec:soundscape_create} for details). Since PlugSonic Soundscape Create allows for playback and navigation, at any moment during the curation, the user can listen to the soundscape and test the exhibition. After this curation steps are completed, the user can save and publish it. At this point, other users could explore the soundscape using their own device (personal computer, mobile or tablet). Some might choose to do so from home, using Soundscape Experience Web to visit the exhibition in the virtual environment. Others might be in the real space (e.g. gallery, museum, archaeological site), or might just decide to explore the soundscape through physical movements within a space, and would therefore choose the Soundscape Experience Mobile app.

During the design process, and whenever a decision needed to be taken about user experience/interface, accessibility and ease of use, the authors intent was to follow the principle stated in the Article 4 of the Convention on the Value of Cultural Heritage for Society – Faro Convention, 2005 \cite{Faro}: “Everyone, alone or collectively, has the right to benefit from the cultural heritage and to contribute towards its enrichment”.
Therefore, differently from most of the tools described in previous paragraphs, the focus of our project on inclusivity and participation required us to develop intuitive and immediate tools, usable by anyone without specific training, and ensuring a true impact on cultural heritage.

\subsection{IMPLEMENTATION: Technologies, Programming Languages and Libraries}

\input{table_apps_libraries.tex}

\begin{figure*} [ht!]
	\centering
	\includegraphics[width=1\linewidth]{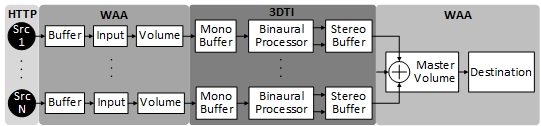}
	\caption{PlugSonic Soundscape signal chain}
  \label{fig:soundscape_signal_chain}
\end{figure*}

This paragraph is devoted to introduce and, where necessary, describe in detail: technologies, programming languages, libraries and APIs used during the development of the \textit{PlugSonic suite}. A complete overview of these is shown in Table \ref{table:technologies} while a description of each app's user interface and features is given in the following sections.
All the web apps in PlugSonic use established web development technologies. HTML and CSS languages are used to create the web page and define its aspect; while JavaScript (JS) to make the apps dynamic and interactive.
The user interface (UI) is built with the React \cite{ReactJS} JS library. The main reason behind the use of React is its efficiency in managing the UI's rendering. With this library, the UI is assembled as a hierarchy of components - for which we define aspect, available interactions and behaviours. Whenever it is necessary to update the UI, e.g. as a consequence of user's actions, instead of re-rendering the whole web page, React will only re-render the components involved. In the case of PlugSonic Soundscape, because of the intensive audio processing associated with each change in the soundscape (e.g. listener's or sources' position change), it was extremely important to have an efficient and smooth UI rendering.
To manage the apps' state and properties, and therefore the apps’ UI and audio processing, we use the Redux API \cite{Redux}. Redux represents the whole app state as a single JS Object, updates the state as the user interacts with the app and triggers the re-rendering process when a state change occurs. Such a centralised approach simplifies the understanding and the update or extension of our web apps with new features - as well as the debug, since it is allows to have an overview of all the properties determining the app's aspect at any given time.
In PlugSonic Soundscape we also adopt the Redux-Saga JS library \cite{ReduxSaga} to manage the side-effects of user's interaction on the sound rendering in an asynchronous fashion. The library allows to define event listeners on the Redux state. When events occur, separate threads execute the code that affects the sound rendering signal chain without interfering with the UI rendering.
In our case, examples of actions that generate side-effects - and either change the audio signal chain structure and/or parameters - are: change listener's/sources' positions, change listener's/sources' settings, add/delete or activate/deactivate sound sources, start/stop playback.
To give Sample and Soundscape a modern and familiar look we adopt the Material design principles \cite{MaterialDesign} through the use of the Material-UI library \cite{MaterialUI}. Material design was formalised by Google with the aim to synthesise "the classic principles of good design with the innovation of technology and science". The Material-UI library is essentially a set of React components that implements those principles; to put it into more tangible terms, these components are the ones used to design every Android mobile operating system and Android-based application and websites many users are already accustomed to. With the same idea of familiarity and immediate understanding, all our web apps use the Material Icons available from Google \cite{MaterialIcons}.
The apps also use responsive design - to scale the UI to different devices and screen sizes - and are \textit{touch enabled}, to allow for their use on touch screens.
The Web Audio API is used to manage the sound processing chains in both PlugSonic Sample and PlugSonic Soundscape. In the Sample app, the WAA is used to implement all the filters and effects (using the \textit{BiquadFilterNode}, \textit{DynamicsCompressorNode} and \textit{ConvolverNode} classes) as well as the audio export - which is obtained by rendering the signal chain in a memory buffer - ready to be downloaded as an audio file. The WAA is also used within wavesurfer.js \cite{Wavesurfer}, which handles the audio playback and waveform visualisation in the Sample app.
Within Soundscape (see Figure \ref{fig:soundscape_signal_chain}), the WAA is only used to manage the connections between sound sources, 3d audio rendering engine and output using \textit{AudioNode} objects.
For each sound source two objects are created: an input node, connected to an input buffer to read the sound samples; and a gain node, to control the volume.

Each gain node is then connected to the mono input buffer of a binaural rendering processor.
The spatialised stereo output is then connected to the master volume summing and gain node.
The master volume gain node is finally connected to the \textit{AudioContext} destination node which represents the actual audio-rendering device (speakers or headphones). The binaural rendering processor can also be bypassed since the users have the option to disable spatialisation for any sound source.
Soundscape also allows to record the binaural signal in real-time and save it as a stereo wav file using the RecorderJS plugin \cite{RecorderJS}.
For the binaural processing we use the 3D Tune-In Toolkit \cite{3DTIToolkit}, an open-source spatialisation library developed during the 3D Tune-In EU project.
The library is presented in \cite{Cuevas2017:Open} and gives control over full 3D listener's and sources' position and movement. With respect to the WAA the 3D Tune-In Toolkit allows for full customisation in terms of HRIRs selection, HRIRs length and sampling rate. Also, the interaural time difference (ITD) is computed from the listener’s head circumference and the head shadowing effect on the interaural level difference (ILD) is simulated for near-field sound sources.
The library also implements a high-performance mode which adopts an IIR filter approximation of the HRIRs for a less demanding, but also less realistic, spatial processing.
A further advantage is obtained in the distance simulation. The library gives the option to choose the amount of attenuation with distance in dB per metre and simulates far-field sources using a low pass filter to emulate air absorption.
Being natively written in C++, we use a JS compiled version \cite{3DTIToolkitJS} generated using the \textit{emscripten toolchain} \cite{Emscripten}.
The exchange of information between the Soundscape web apps (Create and Experience) and the Soundscape Experience mobile app is possible using a JS Object Notation (JSON) file which contains all the relevant data about the soundscape (e.g. sources' position and settings, room size and shape, room floor plan image, URL links to the sound files etc.).

\subsection{PLUGSONIC SAMPLE}

As explained in the introduction, PlugSonic Sample can be used to edit and apply audio effects to any audio file uploaded to the PLUGGY social platform. The modified file can then be saved into the social platform and used in any exhibition (e.g. Blog Story, Time Line, PlugSonic Soundscape).

\begin{figure*}[ht!]
	\centering
	\includegraphics[width=1\linewidth]{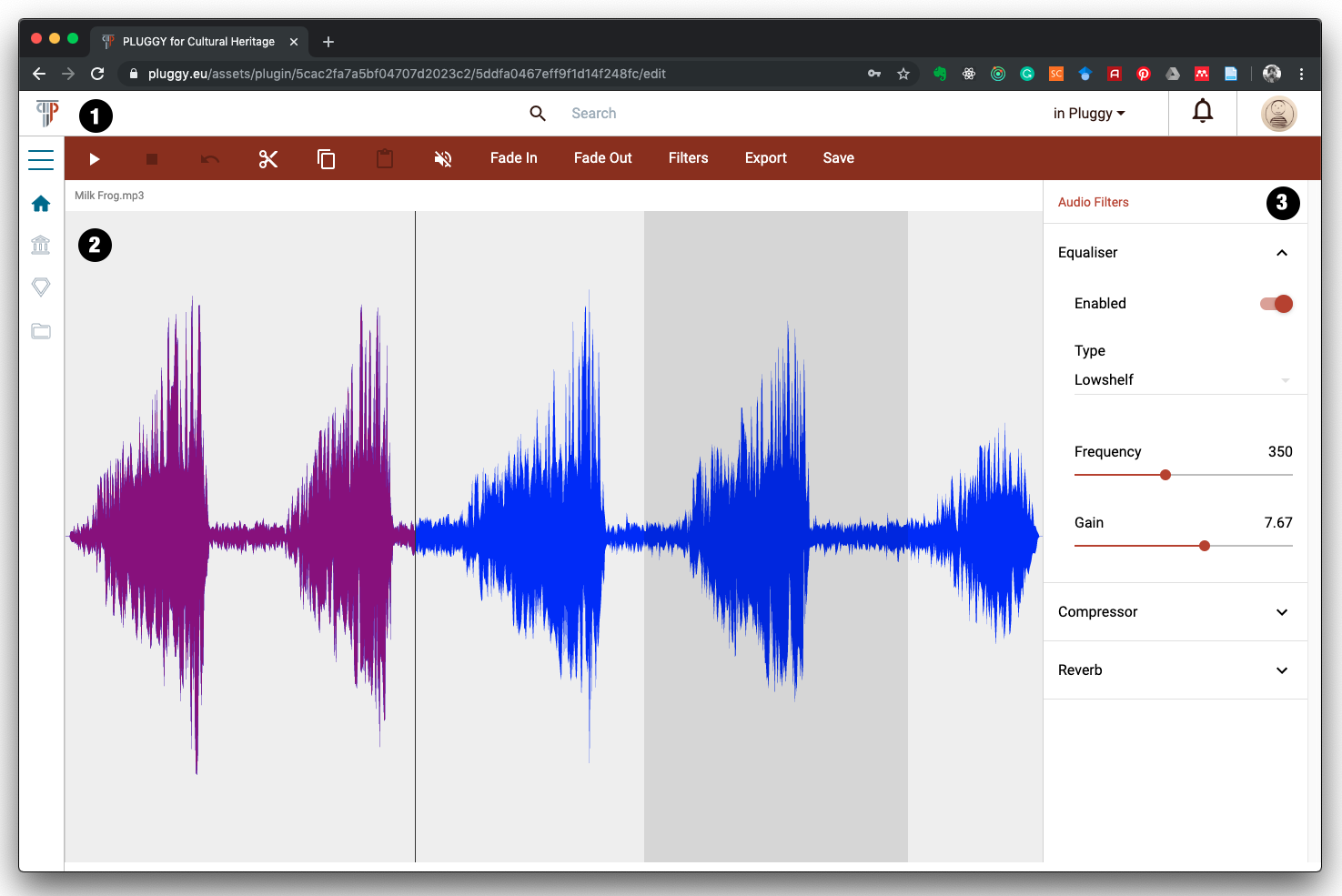}
	\caption{PlugSonic Sample User Interface}
    \label{fig:UI_Sample}
\end{figure*}

Figure \ref{fig:UI_Sample} shows the UI of the web application, integrated into the PLUGGY social platform. The UI is divided into three main parts: (1) playback and edit controls, together with buttons to export the audio file and save the modifications; (2) waveform display with mouse navigation and selection functions; (3) filters and effects menu. For a complete description of the controls and features implemented in Sample we refer to the Appendix A.

\subsection{PLUGSONIC SOUNDSCAPE CREATE} \label{sec:soundscape_create}

As illustrated in the previous sections, PlugSonic Soundscape Create was developed for the curation of interactive 3D audio narratives and soundscapes. To do so, a user proceeds with the creation of a new exhibition through the PLUGGY social platform. After selecting \textit{Soundscape} as the exhibition's type and setting its title and description the user is presented with the Soundscape Create UI, allowing them to proceed with the curation.

\begin{figure}[ht!]
	\centering
	\includegraphics[width=1\linewidth]{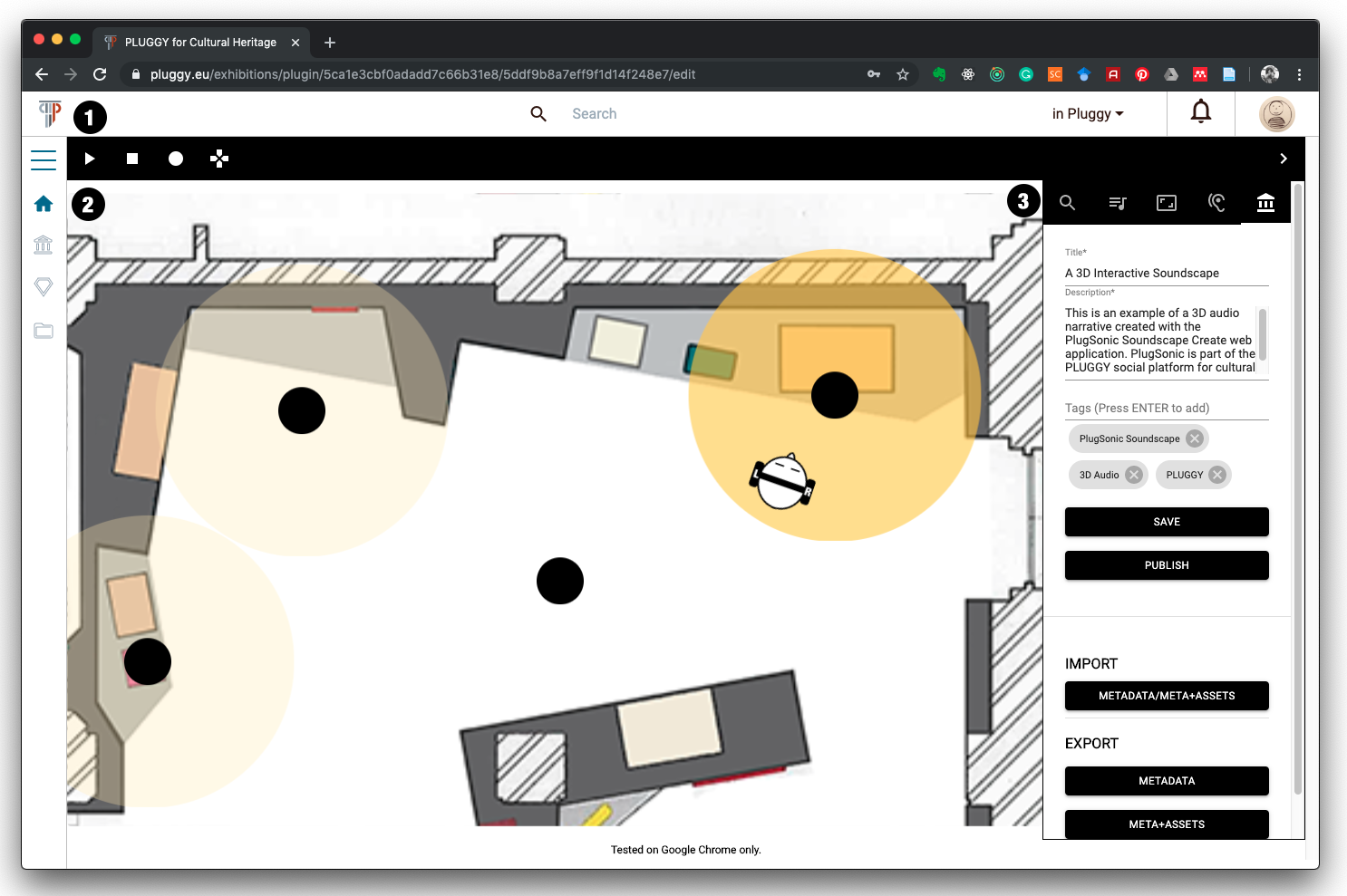}
	\caption{PlugSonic Soundscape Create User Interface}
    \label{fig:UI_Soundscape}
\end{figure}

The UI, shown in Figure \ref{fig:UI_Soundscape}, is divided into three main sections: (1) the top bar, hosting playback control buttons; (2) the "room", which shows the physical/virtual space described in the soundscape and includes icons that represent sound sources and listener; (3) the dismissable side menu, containing all the controls and options to modify the soundscape.

\begin{figure}[ht!]
	\centering
	\includegraphics[width=1\linewidth]{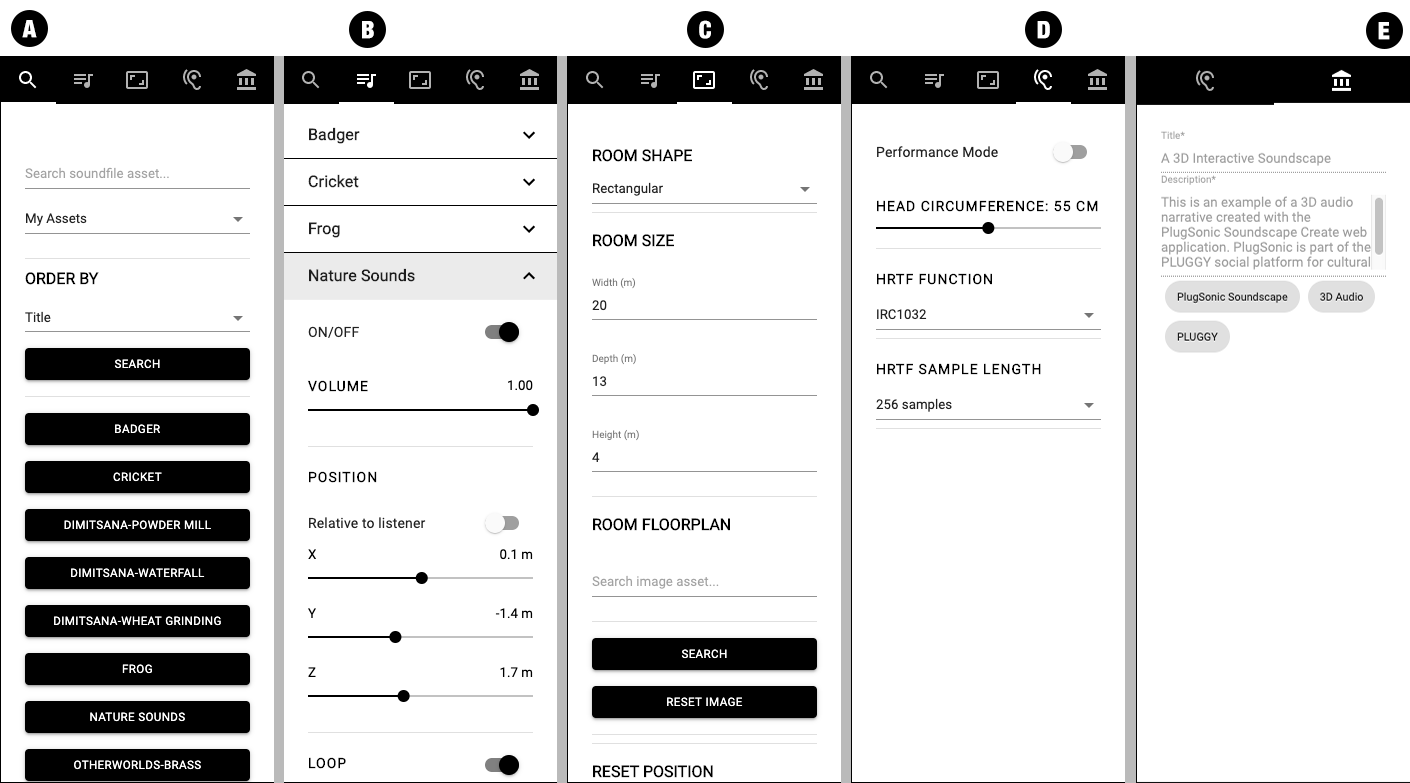}
	\caption{PlugSonic Soundscape Dismissable Menus}
    \label{fig:UI_Soundscape_MenuAll}
\end{figure}

The curation process would typically proceed as follows. The user starts by setting shape (round/rectangular) and size (width/depth/height) of the room and, if desired, choose an image to be used as floor-plan (\textit{room} tab - Figure \ref{fig:UI_Soundscape_MenuAll}.C). After uploading the sound sources - using the \textit{search} tab (Figure \ref{fig:UI_Soundscape_MenuAll}.A) - the user sets the options for each sound source (\textit{sound sources} tab - Figure \ref{fig:UI_Soundscape_MenuAll}.B). Appendix B includes a complete description of all the options available for each sound source, here we limit our description to what we believe are the most interesting and useful controls: \textit{Position} - absolute or relative to the listener; \textit{Loop} - to choose if the sound source will loop or play only once; \textit{Reach} - to control the interaction between listener and sound source (when ON the listener will be able to hear the sound source only when inside the interaction area); \textit{Reach radius} - to set the size of the interaction area; \textit{Timings} - to set an order in the reproduction of any sound source by constraining the playback to the reproduction of another sound source.

At any point during the soundscape’s creation, the user can explore the soundscape to verify the results. It is also possible to take a recording while navigating the soundscape in real-time and export it as a .wav audio file, which will include all the properties of the 3D audio rendering in a standard stereo audio file.

The exhibition's properties (Figure \ref{fig:UI_Soundscape}) can be modified if necessary and the soundscape saved to the social platform and published. There are also buttons available to export the soundscape’s metadata. This is to allow the user to keep a local copy of the soundscape. Also, the metadata can be imported into any application capable of interpreting it. In the case of PlugSonic, the metadata can be imported back into Soundscape Create or it can be opened with Soundscape Experience Mobile (see section \ref{sec:soundscape_experience_mobile}). There are two formats available to export the soundscape. The simple metadata, which requires an internet connection to access PLUGGY's social platform and retrieve the audio files; or the metadata including the assets, in which case, the soundscape can be experienced off-line since the audio files' data is added to the exported soundscape file.

The UI presented here is the result of both: a complete redesign and extension, as well as the outcome of the expert's evaluation described in \cite{Comunita2019:Web}. For a complete description of the controls and features available in Plugsonic Soundscape Create we refer to Appendix B.

\subsection{PLUGSONIC SOUNDSCAPE EXPERIENCE WEB}
The Soundscape Experience web app was developed to allow the navigation of \textit{Soundscape} exhibitions using any device capable of running a web browser (pc, laptop, tablet or mobile). The app's UI is the same as Soundscape Create stripped down of all the features that allow to modify the soundscape - the only controls available are the playback and record buttons, the access to the touch arrows controls, and the listener’s options (Figure \ref{fig:UI_Soundscape_MenuAll}.D). Also, the exhibition’s title, description and tags are visible but not editable (Figure \ref{fig:UI_Soundscape_MenuAll}.E). Soundscape Experience is loaded when a user clicks on the \textit{view} button on the exhibition's page within the social platform.

\subsection{PLUGSONIC SOUNDSCAPE EXPERIENCE MOBILE} \label{sec:soundscape_experience_mobile}

The Soundscape Experience Mobile application has been designed with two main goals: first, to enable navigating soundscapes using a natural, touch-based interface at home or on the go. Second, to allow users to explore soundscapes in an immersive virtual experience delivered in real-world environments. These goals are accomplished by providing two separate interaction modes.

\begin{figure}[ht!]
	\centering
	\includegraphics[width=.7\linewidth]{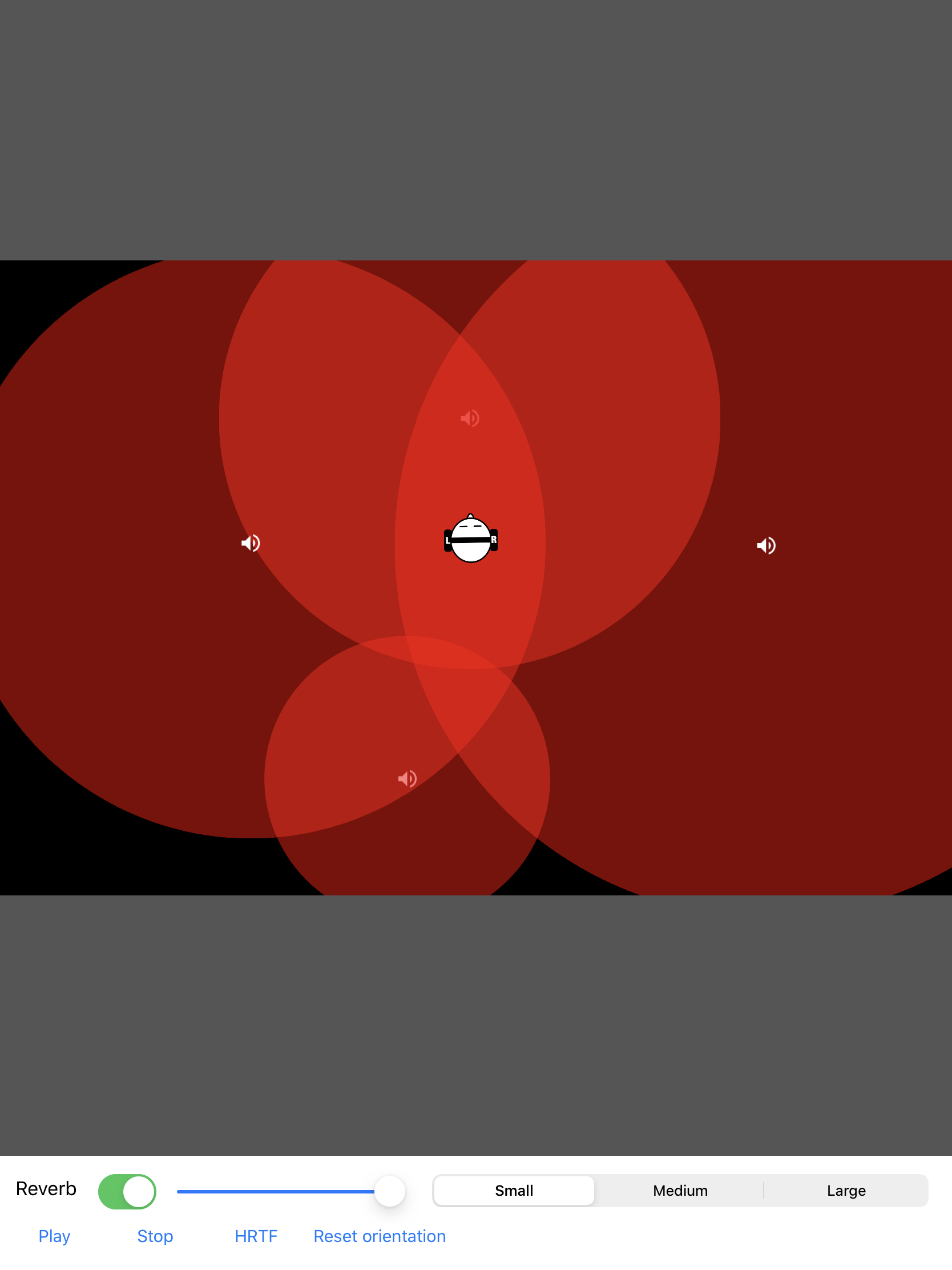}
	\caption{PlugSonic Soundscape Experience Mobile - touch-based UI}
    \label{fig:UI_Soundscape_Mobile01}
\end{figure}

In the first mode (Figure \ref{fig:UI_Soundscape_Mobile01}), users can explore the Soundscape by moving the listener's icon using their finger and can change orientation by rotating the device. The 3D audio simulation is updated in real-time according to the listener's icon current position.

\begin{figure*} [ht!]
	\centering
	\includegraphics[width=.7\linewidth]{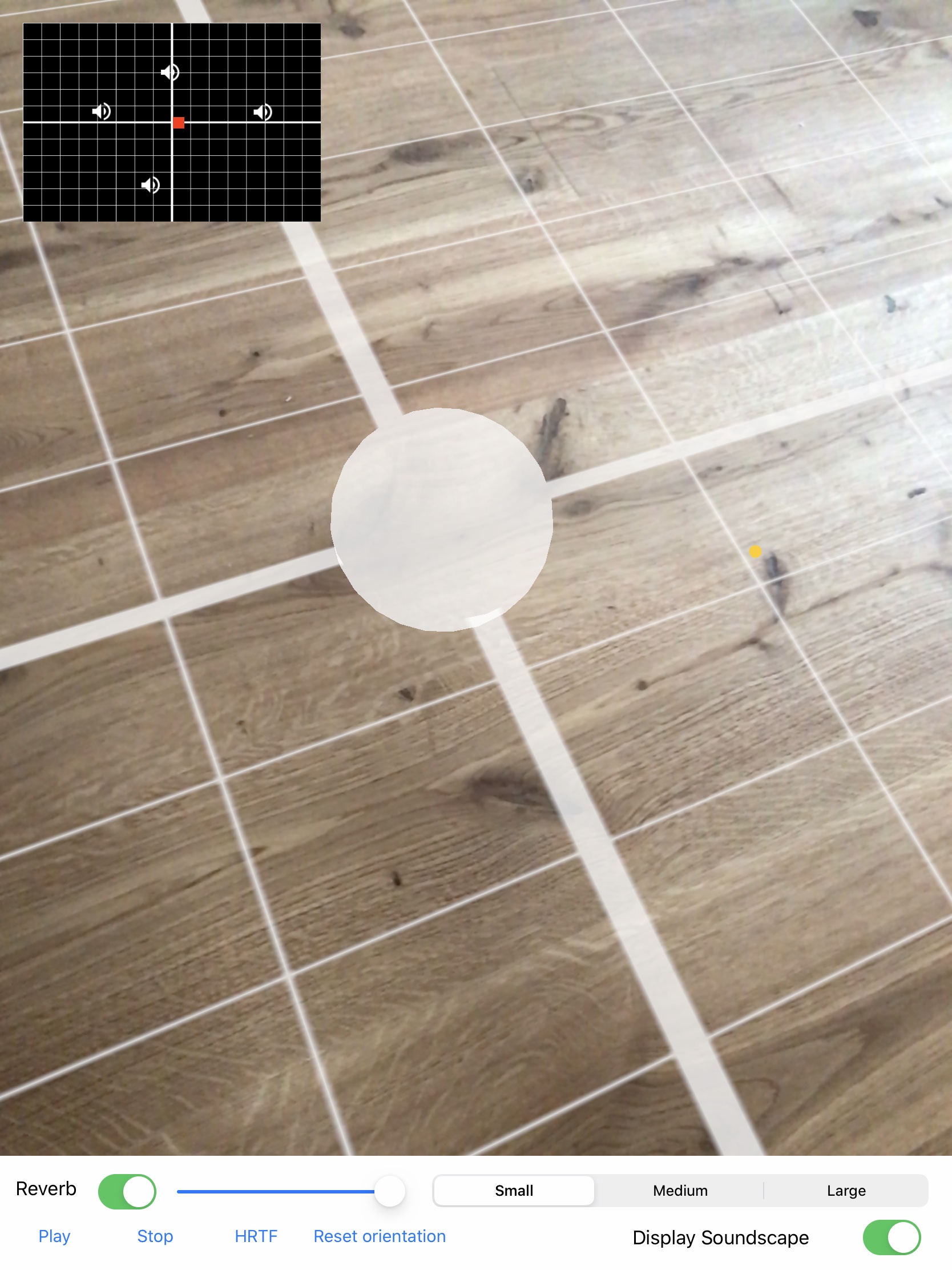}
	\caption{PlugSonic Soundscape Experience Mobile - camera- and sensors-based UI}
    \label{fig:UI_Soundscape_Mobile02}
\end{figure*}

The second interaction mode aims to provide an immersive experience by enabling users to explore a soundscape according to their movements in the real-world. In order to achieve such goal, we use ARkit \cite{ARKit}, a technology developed by Apple to easily support detecting and tracking planar surfaces by analysing video frames captured by a device's camera and data collected by inertial sensors. The framework provides anchors in a real-world environment that can be used to determine the relative position of the device with respect to the detected plane. Such position is used to update the 3D audio simulation in real-time (Figure \ref{fig:UI_Soundscape_Mobile02}).

Soundscapes can be loaded from the PLUGGY social platform using the QR code reader included in the app or importing the metadata. The app has also controls for the reverb settings, a button to reset the listener’s orientation, play and stop button, and a button to choose the HRTFs.

\section{EVALUATION}
The evaluation of PlugSonic focused on the Soundscape Create and Soundscape Experience Mobile apps, which we considered the most critical in terms of contributions to both the spatial audio and cultural heritage research communities.
The purpose of the evaluation was manifold. Understand whether users with no previous experience in sound design or cultural heritage content curation, and without previous knowledge and experience of 3D audio could easily familiarise with the apps and use the functionalities and features as they were designed for. See if users could quickly and effectively use PlugSonic to recreate a 3D soundscape that maps onto a real physical space. Find out if spatial audio technologies in general and the Experience Mobile app in particular constitute a useful and practical way to design compelling experiences and improve engagement with and understanding of cultural heritage. Specifically, we intended to answer the following questions:
\begin{itemize}
    \item How would users create a soundscape using PlugSonic?
    \item Could users easily find functionalities/features they need?
    \item Did users understand all functionalities/features offered in Plugsonic?
    \item If not, what issues did participants face?
    \item How long did it take them to successfully use the functionality/feature?
    \item How easy or difficult was it for users to recreate a 3D soundscape experienced in a real physical space?
\end{itemize}

The evaluation was done in 3 parts. In the first part, participants created a soundscape following predefined tasks provided to them. In the second part, the same participants listened to a soundscape set up in a real physical space and were then asked to recreate it using Soundscape Create. In the third part, another group of participants was asked to explore the soundscape created for part 2 using Soundscape Experience Mobile. We discuss the details further below.\\

\subsection{PART 1 - SOUNDSCAPE CREATION}

\subsubsection{Part 1 - Methodology}
\begin{figure}
	\centering
	\includegraphics[width=.5\linewidth]{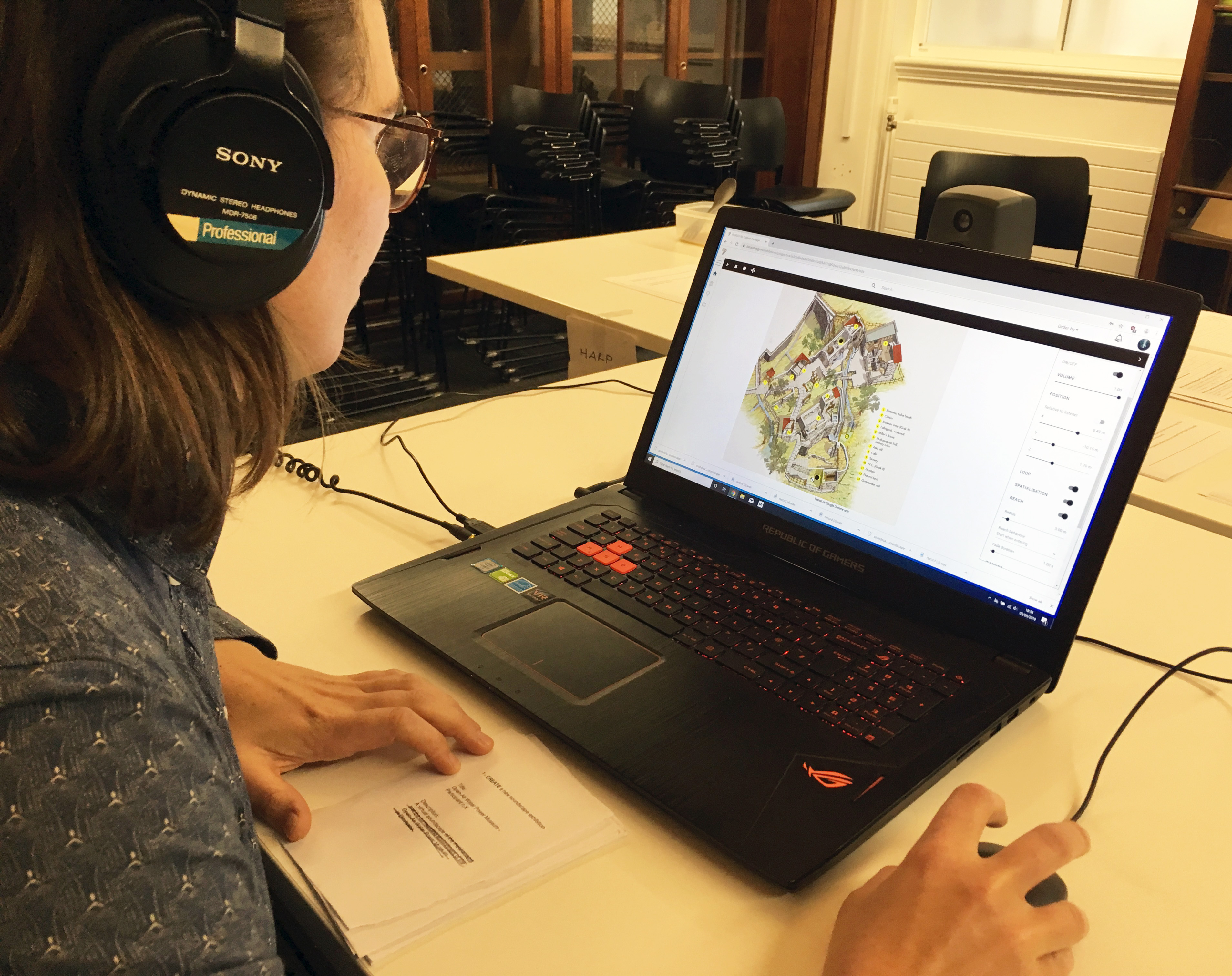}
	\caption{Participant creating a soundscape for part 1 of the evaluation}
    \label{fig:eval_part1}
\end{figure}

In the first part, participants were asked to create a soundscape for one of the institutions involved in the project, the Open-Air Water Power Museum in Dimitsana (Greece) . They were asked to use the Soundscape Create web app together with the material (images and sounds) already available on the PLUGGY social platform. Initially, participants were given a verbal introduction to the PLUGGY project and shown the PLUGGY social platform website, how it is organised and how to navigate it. They were then given 5 minutes to get familiar with the application, explore the different features, and ask questions for clarifications whenever necessary. Participants were then given 12 tasks (Table \ref{table:eval_part1}) that would lead to a complete soundscape exhibition (Figure \ref{fig:eval_part1}). Participants were asked to think aloud and, after each task, to rate how easy it was on a 7-point Likert scale - with 1 as extremely difficult and 7 as extremely easy (Single Ease Questions, SEQ). Time required to complete each task was measured and observations were taken throughout the experiment. We recruited 5 participants for part 1: 2 males and 3 females young academics in their mid 20s to mid 30s. All participants reported to have normal hearing and no previous experience with 3D soundscapes.

\subsubsection{Part 1 - Results}
Overall, most of the steps in task 1 were easy for participants to complete and the average rate was 6.4 (out of 7). See table \ref{table:eval_part1} for the average ease score and completion time per task.

However, participants faced some difficulties for 3 sub-tasks. For example, users had difficulties adding a new sound source. One participant did not use the \textit{My Assets} option for the search. One participant exited the app to search in the "My Assets" page of the social platform and then tried the \textit{Import} button meant to be used for soundscape metadata (see section \ref{sec:soundscape_create}). Once assets were found, 2 participants tried to drag and drop the sound source button from the search panel onto the room instead of clicking on it. Once an asset was added, one participant could not see the icon of the source as it was masked by the floor-plan image. Participants also had difficulties understanding the hinting system, as it was perceived as not giving the right suggestions based on the word being typed. Two participants could not set up the reach of the sound and 2 other participants had issues with the slider precision.

We also observed difficulties when users were asked to set up the sound sources so that it would start playing when the user enters the area of reach. One participant struggled to find the drop-down menu and it was not clear to them what ''start when entering'' meant. One participant had difficulty seeing when a new sound source panel started and commented it was too easy to accidentally click on the delete button.

Finally, issues were observed when adding tags to the exhibition. One participant could not find the \textit{Exhibition} tab, whereas another struggled to find the \textit{Tags} field and checked throughout all tabs. Once found, one participant did not press Enter to add the tag as the call to action was not clear.

\begin{table}
    \centering
    \begin{tabular}{llll}
        \hline\hline
        & Task description & Average & Average\\
         & &  SEQ &completion\\
          & &  & time (sec)\\
        \hline
        1   & CREATE a new soundscape exhibition            &  6.8 & 73 \\
        2   & Set the ROOM SIZE                             &  7 & 16 \\
        3   & Set the ROOM FLOORPLAN image                  &  6.6 & 20\\
        4   & ADD a new sound source                        &   & \\
            &  (search, position and reach)                 &  4.2 & 154\\
        5   & ADD a new sound source                        &  &  \\
            &  (search, position, reach and hidden)         &  6.6 & 62 \\
        6   & ADD a new sound source                        &   &  \\
            &  (search, position, and fade)                 &  6.6 & 82 \\
        7   & Set all sound sources to START                &   & \\
            & PLAYING WHEN ENTERING  reach                  &  5.6 & 70\\
        8   & Set sound sources TIMINGS to play in order    &  6.2 & 74\\
        9   & PLAY the soundscape and MOVE the              &   & \\
            &  listener around to listen to the soundscape  &  7 & 12\\
        10  & Add TAGS to the exhibition                    &  5.8 & 50\\
        11  & SAVE and PUBLISH the exhibition               &  7 & 18\\
        12  & EXPORT the metadata to the laptop             &  6.8 & 14\\
        \hline
    \end{tabular}
    
    \vspace{10pt}
    \caption{Evaluation: Part 1 - Single Ease Questions scores}
    \label{table:eval_part1}
\end{table}

\subsection{PART 2 - SOUNDSCAPE CURATION}

\subsubsection{Part 2 - Methodology}
In the second part participants were asked to become curators, and use what they had learnt about the Soundscape Create app in part 1 to recreate, as faithfully as possible, a virtual soundscape from a real one. For this purpose, we set up a room (rectangular - 14x16x6m) to look like it was part of an exhibition about space exploration. Specifically, we prepared an installation about the Voyager space program. We collaborated with a professional musician who composed and recorded a 5 minutes piece of music to use as accompaniment to speech excerpts from the \textit{Golden Record} \cite{GoldenRecord}. The content was mixed into 6 tracks, each of which was assigned to 1 of 6 loudspeakers placed around the room. A MaxMSP patch was used to control two laptops (a master and a slave) each one driving 3 loudspeakers. The patch allowed to assign tracks to speakers, set the volume of each track, control the playback - and used the Open Sound Control (OSC) protocol for communication between master and slave laptop.

\begin{figure}
	\centering
	\includegraphics[width=.5\linewidth]{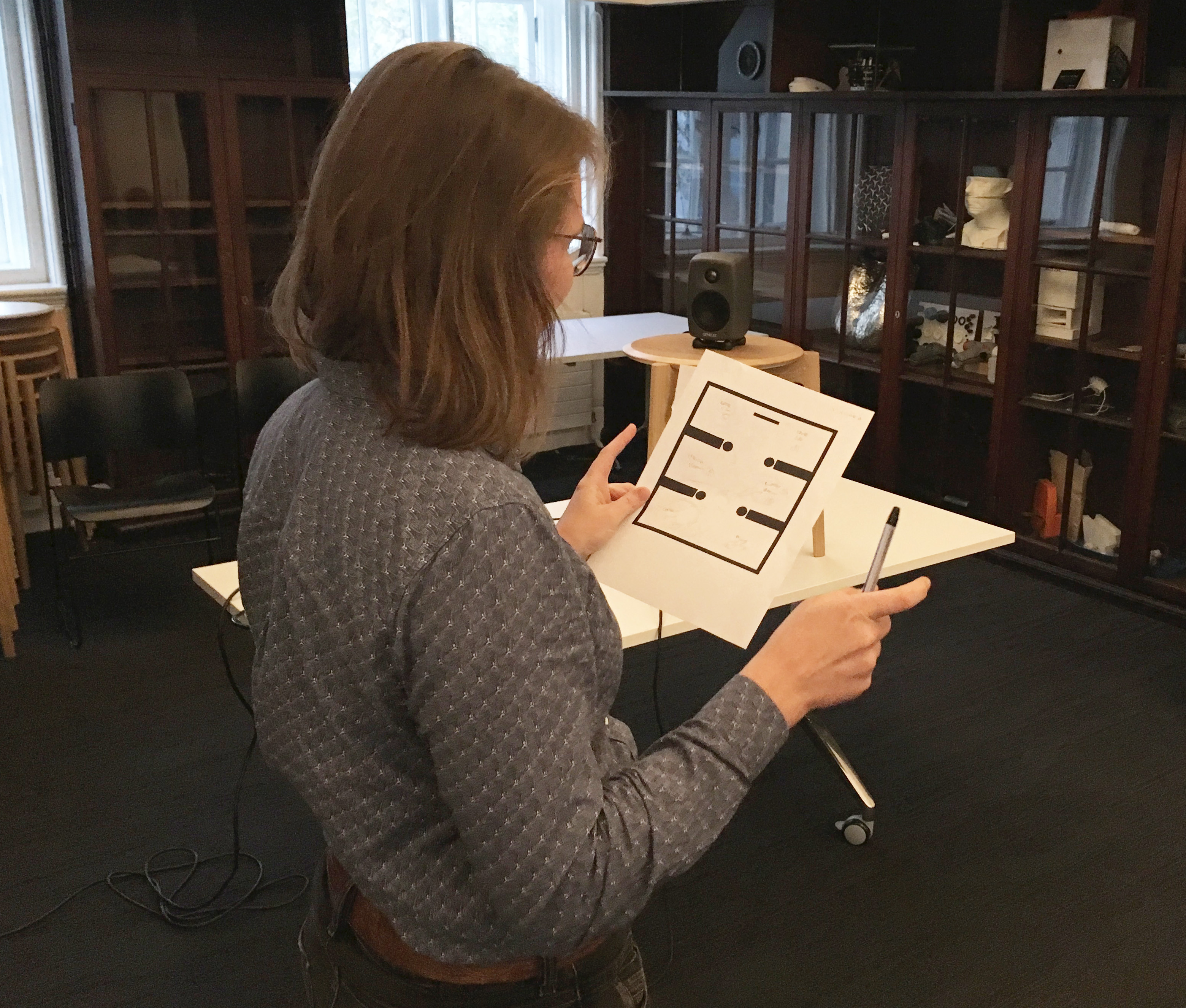}
	\caption{Participant taking notes for the soundscape creation task in part 2 of the evaluation}
    \label{fig:eval_part2}
\end{figure}

Participants were first asked to listen and observe the real soundscape; they were also given a floor-plan of the room which they could use to take notes about the setup (Figure \ref{fig:eval_part2}). Participants were asked to pay attention to the following: (1) match the track to the loudspeaker (each loudspeaker was labelled with the track assigned to it), (2) the loudspeakers’ position (e.g. position in the room, height), and (3) the sound sources’ reach (i.e. from how far the sound of each loudspeaker could be heard). Participants were allowed to listen to the soundscape as many times as needed, and after having noted down all the information, they needed to recreate the soundscape using the material (images and sounds) already available on the PLUGGY social platform. After completing the task, participants were asked to fill in the System Usability Scale (SUS) questionnaire (Table \ref{table:eval_part2}), to measure usability, and score how likely they would recommend PlugSonic to others (net promoter score, NPS). We recruited 5 participants for part 2 (the same who participated in part 1).

\subsubsection{Part 2 - Results}
Overall, participants agreed that the application is easy to use and were confident in using it, but there is space for improvement with regards to clarity of features, iconography, and the integration of features.

Specifically, participants found the visual and interactive aspect of PlugSonic most interesting as it gives quick results and a pleasant user experience. They also liked the possibility to test the soundscape and listen to the binaural audio directly on the Create app. What participants did not like about the application is that it first requires some level of understanding of the available features and icons. Furthermore, some tasks where judged as being "too manual" and requiring "too many clicks" to implement. Additional features were desirable such as being able to save the settings of a single sound and automatically close the source panel or choose the colour for the sources' icons.

Participants were neutral in whether they would use the system frequently. This is consistent with their rating of a 6 (out of 10) to the question how likely they would use any of the curation tools and add stories to the social platform. Although they see the novelty in the interactivity, the main reason for these scores is the fact that they do not work in the sector specifically or are not used to creating content.

For the NPS score: among the participants we had 1 detractor (one participant rated a 6), 2 passives (2 participants rated an 8) and 2 promoters (2 participants rated a 9). This results in an overall NPS score of 20, which is good. The list of questions with the average rating is shown in Table \ref{table:eval_part2}

\begin{table}
    \centering
    \begin{tabular}{llll}
        \hline\hline
        & Question & SUS \\
        \hline
        1   & I think that I would like to use this system frequently.  &  4.4 \\
        2   & I found the system unnecessarily complex.                 &  2.8 \\
        3   & I thought the system was easy to use.                     & 6.2 \\
        4   & I think that I would need the support of a technical      &   \\
            &  person to be able to use this.                           &  3.0  \\
        5   & I found the various functions in this system were         & \\
            & well integrated.                                          & 5.4 \\
        6   & I found functionality of features and controls clear.     &  5.0  \\
        7   & I thought there was too much inconsistency                &  \\
            & in this application.                                      &  1.6  \\
        8   & I would imagine most people would learn to use            &   \\
            & the application very quickly.                             & 5.2  \\
        9   & I felt very confident using the system.                   &  5.8  \\
        10  & I needed to learn a lot of things before I could          & \\
            &  get going with this application.                         & 1.6 \\
        \hline
    \end{tabular}
    
    \vspace{10pt}
    \caption{Average system usability score for part 2 of the evaluation}
    \label{table:eval_part2}
\end{table}

\subsection{PART 3 - SOUNDSCAPE EXPERIENCE}

\subsubsection{Part 3 - Methodology}
\begin{figure}
	\centering
	\includegraphics[width=.5\linewidth]{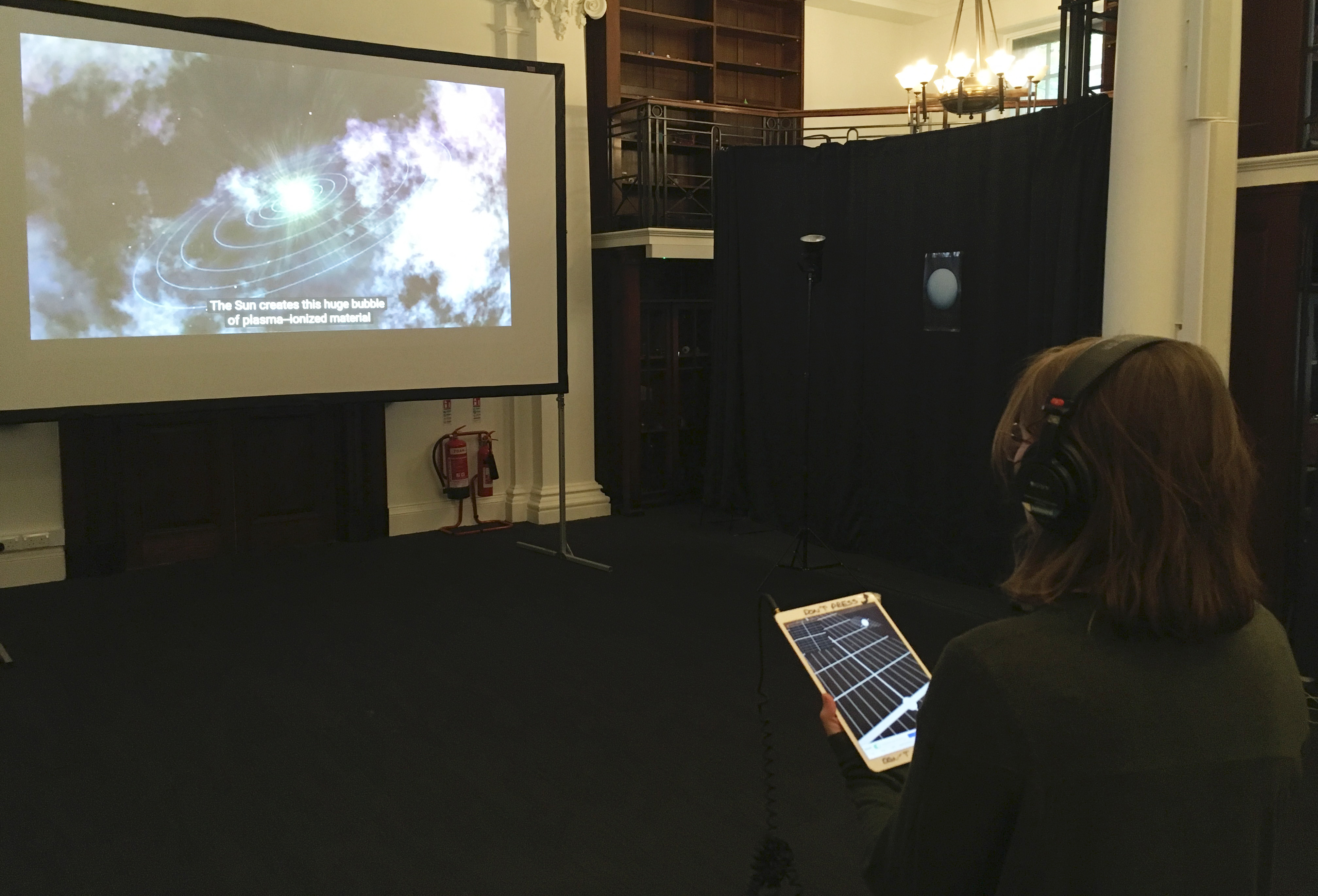}
	\caption{Participant using the Soundscape Mobile app to explore the soundscape in a real space}
    \label{fig:eval_part3}
\end{figure}

In part 3, participants were asked to use the PlugSonic Experience Mobile app to explore a soundscape within a real environment. The soundscape and physical space were the same used for part 2. The room was set up with photos and a video projection about the Voyager exploration program (Figure \ref{fig:eval_part3}).
Participants were asked to imagine visiting a museum and as part of the visit they were given a mobile device (iPad) which they could use to explore a soundscape. They were invited to explore the soundscape freely for as long as they wished while moving around the room. No specific recommendations were provided apart from the indication to point the device's camera in the direction they were facing - since the camera and inertial sensors were used by the device to infer position and orientation, which in turn were used by the mobile app to render the audio in 3D. They were also advised to ignore the device's screen since no information was going to be displayed. After exploring the soundscape, participants were asked to fill in a questionnaire about their emotional response, social potential, and learning while using the app. We recruited 7 participants for part 3: 5 males, 2 females in their mid 20s to mid 30s.

\subsubsection{Part 3 - Results}
During this experience, most participants have indicated feeling "interested" (7 participants), "engaged" (6 participants), and "captivated" (6 participants). Some indicated to have felt "excited" (2 participants), "indifferent", "satisfied", "inspired", "immersed", "calm", "overwhelmed", or "confused". The average score for personal resonance and emotional connection was 5.2 out of 7 (see table \ref{table:eval_part3}a). The average score for learning and intellectual stimulation was 4.4 (see table \ref{table:eval_part3}b). The average score for shared experience and social connectedness was 5.8 (see table \ref{table:eval_part3}c). This means future improvements should be made to increase scores for personal resonance and emotional connection and learning and intellectual stimulation. It is important to underline how the soundscape adopted for the evaluation did not contain any speech or narration describing the material presented, which might help to explain the score for learning and intellectual stimulation.

\begin{table}
    \centering
    \caption{Evaluation part 3}
    \label{table:eval_part3}

    \begin{tabular}{p{0.03\textwidth}p{0.7\textwidth}ll}
        \hline\hline
        & Questions  & Likert  \\
        & \textbf{Personal resonance and emotional connection} & Score \\
        \hline
        1   & The soundscape made experiencing the objects      &   \\
            &  more interesting/fun.                            &  5.4 \\
        2   & I found the soundscape emotionally engaging.      &  5.6 \\
        3   & During the experience, I felt connected with the  & \\
            &  objects presented to me                          & 4.1 \\
        4   & I will be thinking about this experience          &  \\
            &  for some time to come.                           &  5.7 \\
        \hline
    \end{tabular}
    \subcaption{}
    \bigskip

    \begin{tabular}{p{0.03\textwidth}p{0.7\textwidth}lp{0.01\textwidth}}
        \hline\hline
        & \textbf{Learning and intellectual stimulation} & &\\
        \hline
        5   & I got a good understanding about the              &   \\
            &  objects presented to me.                         &  3.1 \\
        6   & I got a good understanding about where            & \\
            & the objects where located.                        & 4.9 \\
        7   & I felt engaged with the objects presented to me.  & 5.0 \\
        8   & I felt challenged and provoked.                   & 4.0 \\
        9   & During this experience, my eyes were opened       &   \\
            &  to new ideas.                                    &  4.9 \\
        \hline
    \end{tabular}
    \subcaption{}
    \bigskip

    \begin{tabular}{p{0.03\textwidth}p{0.7\textwidth}lp{0.01\textwidth}}
        \hline\hline
        & \textbf{Shared experience and social connectedness} & &\\
        \hline
        10  & I would have liked to have shared about           &  \\
            & it with other people                              & 5.7 \\
        11  & After this experience, I wanted to talk           &   \\
            &  to people about it.                              &  5.9 \\
        \hline
    \end{tabular}
    \subcaption{}
\end{table}

\subsection{Discussion}
The evaluation aimed at understanding if PlugSonic (1) is easy to learn and use for people unfamiliar with 3D audio in general and the concept of soundscape in particular; (2) is an effective tool for the creation of 3D soundscape from and for real environments; (3) has the potential to improve engagement and understanding of cultural heritage.

Overall, creating a soundscape was rated with a 6.4 (out of 7) for ease. Participants felt confident using PlugSonic Soundscape and found the interactive aspect of creating the soundscape most interesting and engaging. Some difficulties were observed, indicating that improvements could be made for adding and setting up new sound sources and tagging exhibitions. More in general, further improvements are possible to make the apps more streamlined and speed up the creation process. At the time of writing, the authors already updated PlugSonic Create based on the feedback obtained during the evaluation. For example, the reach radius slider visibility has been improved; a visual feedback has been added when listener is in reach of a source (reach area changes colour); precision for all sliders has been set to one decimal point; the separation between sources' panels has been made clearer; and a "Loading.." message has been added when retrieving a sound source.

With a net promoter score of 20, participants are likely to recommend PlugSonic Soundscape to others. Although, one of the most critical aspects highlighted during evaluation was the resistance of users to become content creators. It seems clear that efforts are still necessary from research communities and institutions to empower citizens and allow them a more active participation in the act of sharing and protecting cultural heritage.

Additionally, for the experience part of PlugSonic Soundscape, positive results were obtained in terms of personal resonance as well as shared experience and social connectedness. Future improvements should be made to support emotional connection and intellectual stimulation better. Although, it needs to be taken into account that the evaluation did not take place in an actual museum with real artifacts being exhibited.

We can conclude that, even accounting for its current limitations, the \textit{PlugSonic suite} represents a major contribution to the digital heritage community, including both end users and researchers. Our evaluation with inexperienced participants showed that the apps and the creation, curation and experience processes, within the support of the PLUGGY social platform, were easily understood and effectively utilised. The speed with which participants were able to familiarise and perform tasks with PlugSonic Create shows how the design and implementation choices seem to have worked to simplify the learning process. Participants appreciated the aspect and the interaction with the app and did not feel limited during the recreation of a virtual soundscape from a real one. This was a particularly important aspect for the project, proving that 3D audio technologies can indeed be democratised without sacrificing the rendering quality and interaction flexibility. Also the Experience Mobile app was well received, there was no particular friction in the navigation of the soundscape within a real space. From the questionnaires' answers it seems clear how 3D audio soundscapes in general and PlugSonic in particular, can help cultural institutions in their mission to deliver engaging experiences and connect the public with cultural heritage.

\section{CONCLUSIONS and FUTURE WORK}

This paper presents the design, development and evaluation of a series of web and mobile applications - the \textit{PlugSonic suite} - implemented as part of the PLUGGY project. The umbrella project includes the development of a social platform and several apps (AR, 3D Audio, Geolocation, Gamification) to provide users with the necessary tools to shape cultural heritage; both as curators and visitors of virtual or augmented exhibitions.

In section \ref{background}, we described the state of the art for binaural audio, web-based spatial audio and sonic narratives, with special attention to cultural heritage applications. Even if research in these fields have produced significant advances, no specific tools have been developed to democratise spatial audio technologies and encourage the general public to have an impact on tangible and intangible heritage through these.
PlugSonic aimed to demonstrate that spatial audio technologies and software can be designed to be accessible to anyone, without having to compromise on the binaural rendering quality or the flexibility of interactions available to the content creator, whether institutions or general public.

Web- and mobile- based applications, due to their ubiquitous nature, have the potential to be exploited to ease the curation and experience of 3D interactive soundscapes and audio narratives.
In this work, we used these technologies to develop applications that can be used to edit sound files and create, test and experience such soundscapes in a low friction environment that allows to transition from a web browser to a physical space when desired.

After introducing the technology used for the development we described the features and functionalities included in the apps - which were, partially, the result of a previous work with experts in the fields of cultural heritage and/or audio \cite{Comunita2019:Web}.

To understand whether our aim to design tools which are simple to learn and to use was achieved, we conducted an evaluation with subjects with no previous experience and knowledge about 3D audio or soundscape design. The evaluation included 3 parts which focused on the creation, curation and experience of 3D audio soundscapes with the PlugSonic Create Web and PlugSonic Experience Mobile applications.
Participants were able to learn to use the Soundscape Create app quickly and effectively; they judged the task of creating a soundscape fairly simple and appreciated the level of interactivity and the possibility to test the results during the creation. Good results were obtained also for the task of creating a virtual soundscape starting from a real one, which shows how PlugSonic can indeed be effectively utilised to convey the atmosphere of a "real-world" situation. 

The Experience Mobile app, exploiting the Apple ARkit \cite{ARKit} for the localisation of the user within a soundscape in a physical space, showed the potential of augmented exhibitions in increasing emotional resonance and connectedness with cultural heritage.

The feedback received during evaluation was used to improve the user interface and user experience but further work is necessary to simplify and speed up the soundscape creation process. The Soundscape apps could also be improved by introducing moving sound sources, for an even more dynamic experience, and radiation patterns, for a more realistic simulation of sounds' directivity or to emulate occlusion effects. The Soundscape Create app already allows to choose the shape of the virtual environment, limited to rectangular and round spaces, and to use an image as floor-plan or background of a soundscape. Further improvements would include custom room shapes and the options to set movement contraints.

Cultural institutions have started to adopt social media to promote events and stimulate participation, but cannot rely on a common ground when it comes to communication channels and tools to involve their audience. Being developed within the framework of the PLUGGY social platforms, PlugSonic could help to bridge the gap between general public and cultural institutions in an effort to encourage participation, co-creation and sharing of cultural heritage; especially because PlugSonic does not require the development or installation of software or hardware, and can be used on any device. Anyone could, visiting a museum or a monument, retrieve audio narratives from the social platform servers and experience them straight away.

Furthermore, we think that PlugSonic has the potential to become also a web- and mobile-based 3D audio research evaluation tool. In fact, even if web-based audio evaluation tools are available \cite{Jillings2015:Web}\cite{Kraft2014:Beaqlejs}, none focuses specifically on spatial audio topics (e.g. HRTF selection, HRTF adaptation, speech reception threshold, cocktail party effect). Moreover, the hearing loss and hearing aid simulation algorithms available from the 3D TuneIn toolkit would also allow for hearing impairment specific tests. PlugSonic could be used within a listening test framework that lets researchers easily design different types of online tests (e.g. AB, ABX, MUSHRA). Both the web- and mobile-based apps could be exploited for localisation tests or games within a virtual or physical space.

We conclude adding links to the standalone versions of the web apps, to the code repositories and the PLUGGY social platform:
\begin{itemize}
    \item PlugSonic Sample (standalone version): \url{http://plugsonic.pluggy.eu/sample}
    \item PlugSonic Soundscape Web (standalone version): \url{http://plugsonic.pluggy.eu/soundscape}
    \item Repository \url{https://github.com/lpicinali/PlugSonic-soundscape}
    \item PLUGGY social platform: \url{https://pluggy.eu/}
\end{itemize}

\section{Acknowledgement}
This work was supported by the PLUGGY project (https://www.pluggy-project.eu/), European Union’s Horizon 2020 research and innovation programme under grant agreement No 726765.
\bigskip

\bibliographystyle{unsrt}  
\bibliography{refs}  

\end{document}

%% file: table_apps_libraries.tex
\begin{table}
    \small
    \centering
    \begin{tabular}{ |m{3.1cm}|m{1.4cm}|m{2.3cm}|m{2.3cm}|m{2.3cm}| }
    \hline
        \backslashbox[3.5cm]{Technology}{App} & 
        \textbf{Sample} & 
        \textbf{Soundscape Create} & 
        \textbf{Soundscape Experience Web} & 
        \textbf{Soundscape Experience Mobile} \\
    \hline
        Programming Language(s) & 
        HTML, CSS, JS & 
        HTML, CSS, JS & 
        HTML, CSS, JS & 
        Objective-C \\
    \hline
        React &
        \checkmark &
        \checkmark &
        \checkmark &
        \\
    \hline
        Redux &
        \checkmark &
        \checkmark &
        \checkmark &
        \\
    \hline
        Redux-Saga &
        &
        \checkmark &
        \checkmark &
        \\
    \hline
        Material-UI &
        \checkmark &
        \checkmark &
        \checkmark&
        \\
    \hline
        Material Icons &
        \checkmark &
        \checkmark &
        \checkmark&
        \\
    \hline
        Responsive Design &
        \checkmark &
        \checkmark &
        \checkmark &
        \\
    \hline
        Touch Enabled &
        \checkmark &
        \checkmark &
        \checkmark &
        \checkmark \\
    \hline
        Web Audio API &
        \checkmark &
        \checkmark &
        \checkmark &
        \\
    \hline
        Wavesurfer.js &
        \checkmark &
        &
        &
        \\
    \hline
        Recorderjs &
        &
        \checkmark &
        \checkmark &
        \\
    \hline
        3D TuneIn Toolkit &
        &
        \checkmark &
        \checkmark &
        \checkmark \\
    \hline
        Apple AR Kit &
        &
        &
        &
        \checkmark \\
    \hline
        JSON &
        \checkmark &
        \checkmark &
        \checkmark &
        \checkmark \\
    \hline
    \end{tabular}
    \vspace{10pt}
    \label{table:technologies}
    \caption{Technologies used by PlugSonic}
\end{table}